\documentclass[journal,letterpaper]{IEEEtran}

\usepackage{cite}
\usepackage{amsmath,amssymb,amsfonts}
\usepackage{bm}
\usepackage{amsthm}
\usepackage{bbm}
\usepackage{textcomp}
\usepackage{graphicx}
\usepackage{xcolor}
\newcommand{\Rcolor}{blue!70!black}
\usepackage{algorithmic}
\usepackage{tikz}
\usepackage{balance}

\usepackage{caption}
\usetikzlibrary{shapes, arrows, positioning, calc}
\usepackage{booktabs}
\usepackage{tabularx}

\usepackage{subcaption}

\theoremstyle{plain}
\newtheorem{theorem}{Theorem}
\newtheorem{lemma}{Lemma}

\theoremstyle{definition}
\newtheorem{remark}{Remark}

\usepackage[colorlinks=true,       
            linkcolor=blue,        
            citecolor=blue,        
            urlcolor=blue,         
            bookmarks=true,        
            pdfborder={0 0 0}]{hyperref}

\usepackage[all]{hypcap} 

\title{Exact 3-D Channel Impulse Response Under Uniform Drift
for Absorbing Spherical Receivers}

\author{
    Yen-Chi~Lee,~\IEEEmembership{Member,~IEEE,}~
    Ping-Cheng~Yeh,~\IEEEmembership{Member,~IEEE,}~and~
    Chia-Han~Lee,~\IEEEmembership{Member,~IEEE}
    \thanks{Accepted for publication in \textit{IEEE Communications Letters}, 2026.}
    \thanks{This work was supported by the National Science and Technology Council of Taiwan (NSTC 113-2115-M-008-013-MY3). (Corresponding author: Yen-Chi Lee.)}
    \thanks{Y.-C. Lee is with the Department of Mathematics, National Central University, Taoyuan, Taiwan (e-mail: yclee@math.ncu.edu.tw); P.-C. Yeh is with the Graduate Institute of Communication Engineering, National Taiwan University, Taipei, Taiwan (e-mail: pcyeh@ntu.edu.tw); C.-H. Lee is with the Institute of Communications Engineering, National Yang Ming Chiao Tung University, Hsinchu, Taiwan (e-mail: chiahan@nycu.edu.tw).}
}

\begin{document}

\maketitle


\begin{abstract}
An exact channel impulse response (CIR) for the three-dimensional
point-to-sphere absorbing channel under drift has remained unavailable
due to symmetry breaking.
This letter closes this gap by deriving an exact analytical CIR for a fully absorbing spherical receiver under uniform drift with arbitrary direction.
By formulating the problem in terms of joint first-hitting time–location statistics and applying a Girsanov-based measure change, drift effects are isolated into an explicit multiplicative factor, yielding an exact series representation.
The resulting CIR provides a rigorous reference model and enables efficient, noise-free evaluation of key channel metrics without relying on Monte Carlo simulations.
\end{abstract}



\section{Introduction}

\color{black}
\IEEEPARstart{T}{he} point-to-sphere absorbing channel \cite{Jamali:2018} stands as the canonical reference model for three-dimensional (3-D) molecular communication (MC) systems \cite{Akyildiz:2008,Nakano:2013,Farsad:2016}. In quiescent environments, this geometry allows for compact analytical characterizations of first-hitting statistics by exploiting its inherent radial symmetry \cite{Yilmaz:2014}. However, practical MC scenarios are rarely static; drift arises naturally from background flow \cite{Srinivas:2012,Kadloor:2012}, externally induced bias fields \cite{ma2019electric,chou2021molecular}, or concentration-driven transport.

The introduction of advection---even if spatially uniform---fundamentally breaks the radial symmetry that classical formulations rely upon. Unlike the no-drift case, advection creates a directional bias that couples the first-hitting time with the angular absorption location. This symmetry breaking introduces a level of analytical complexity that precludes the direct extension of radial-only models to flow-perturbed environments.

Consequently, despite its physical relevance, an exact analytical \textit{channel impulse response (CIR)} for the point-to-sphere geometry under arbitrary drift direction has remained unavailable since the seminal work \cite{Yilmaz:2014} of Yilmaz et al. in 2014. Existing results are restricted to zero-drift scenarios \cite{Yilmaz:2014}, specifically aligned drift configurations \cite{Akdeniz:2018}, or reduced-dimensional approximations \cite{Srinivas:2012,Kadloor:2012}, while general settings are typically relegated to Monte Carlo simulations or empirical fitting \cite{Jamali:2018,Ahuja:2022}. This lack of an exact reference model has limited both physical interpretability and systematic channel analysis in drift-dominated regimes.
\color{black}

In this letter, we close this gap by deriving an exact analytical CIR for
an absorbing spherical receiver
in 3-D space under
uniform drift with arbitrary direction.
Rather than modifying existing no-drift expressions, we formulate the
problem at the level of the \emph{joint} first-hitting time and location
statistics.
By applying a measure-change argument based on the Girsanov theorem \cite{oksendal2003stochastic}, the
effect of drift is incorporated as an explicit reweighting of boundary
hitting events, while preserving the analytical structure of the
baseline no-drift solution.

The resulting CIR admits an exact series representation that remains
valid for arbitrary drift orientations and strengths.
This formulation provides a rigorous reference model for drifted
point-to-sphere channels, resolving a long-standing symmetry-breaking
regime that has previously been treated only approximately or through
simulation \cite{Jamali:2018,noel2014optimal,Koo:2016,gursoy2019index}.
Beyond channel characterization, the availability of an exact CIR
enables efficient and noise-free extraction of key system metrics, such
as peak time and peak amplitude \cite{Yilmaz:2014,tepekule2015isi}, which are difficult to obtain reliably
from particle-based Monte Carlo methods.

The main contributions of this letter are summarized as follows:
\begin{itemize}
\item[(i)] An exact analytical series expression \eqref{eq:main_result} for the 3-D
point-to-sphere CIR under uniform drift with arbitrary direction.
\item[(ii)] A measure-change-based framework (see Fig.~\ref{fig:logic_flow}), via the Girsanov theorem,
that isolates drift effects into an explicit multiplicative factor and
admits extension to other molecular channel models.
\end{itemize}

\begin{table}[h]
\caption{Comparison of related analytical CIR results in MC literature. Note that in 1-D space, spherical and planar geometries coincide as the receiver reduces to a point boundary.}
\label{tab:related_works}
\centering
\renewcommand{\arraystretch}{1.3} 
\begin{tabularx}{\columnwidth}{@{} l >{\hsize=0.8\hsize\raggedright\arraybackslash}X >{\hsize=1.2\hsize\raggedright\arraybackslash}X @{}}
\hline\hline
\textbf{Dim.} & \textbf{No Drift} & \textbf{Uniform Drift} \\
\hline
$d=1$
& L\'evy distribution \cite{Kadloor:2012,Srinivas:2012}
& Inverse Gaussian (IG) distribution \cite{Kadloor:2012,Srinivas:2012} \\
\hline
$d=3$ (Planar)
& Exact analytical CIR \cite{lee2016distribution,Lee:2024_TCOMM}
& Exact analytical CIR (arbitrary direction) \cite{lee2016distribution,Lee:2024_TCOMM} \\
\hline
$d=3$ (Spherical)
& Exact analytical CIR \cite{Yilmaz:2014}
& \textbf{No exact analytical CIR in prior work}; \cite{Akdeniz:2018} restricted to aligned drift \\
\hline\hline
\end{tabularx}
\end{table}


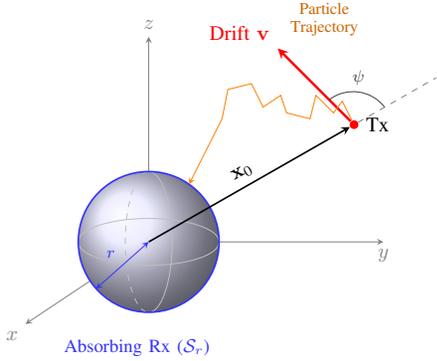
\begin{figure}[t]
    \centering
    \scalebox{0.78}{
    \begin{tikzpicture}[scale=1.0, >=stealth,
        axis/.style={gray, ->, thin},
        vector/.style={thick, ->}]
      
        \coordinate (O) at (0,0); 
        \coordinate (Tx) at (3.5, 2.0); 
        
        \draw[axis] (0,0) -- (-2.1, -1.4) node[anchor=north east] {$x$};
        \draw[axis] (0,0) -- (0, 3.5) node[anchor=south] {$z$};
        \draw[axis] (0,0) -- (4, 0) node[anchor=north] {$y$};
        
        \shade[ball color=blue!10, opacity=0.6] (O) circle (1.2cm);
        \draw[gray!50, thin] (O) ellipse (1.2cm and 0.4cm); 
        \draw[gray!50, thin] (0,-1.2) arc (-90:90:0.4cm and 1.2cm); 
        \draw[gray!50, thin, dashed] (0,1.2) arc (90:270:0.4cm and 1.2cm); 
        \draw[blue!80, thick] (O) circle (1.2cm);
        
        \draw[<->, blue!80] (0,0) -- (-0.9, -0.8) node[midway, above left, font=\footnotesize] {$r$};
        \node[blue!80, font=\small, align=center] at (-0.2, -1.8) {Absorbing \text{Rx} ($\mathcal{S}_r$)};
        
        \node[circle, fill=red, inner sep=1.5pt, label={right:\text{Tx}}] (TxNode) at (Tx) {};
        \draw[vector, black] (O) -- (TxNode) node[midway, above, sloped] {$\mathbf{x}_0$};
        
        \coordinate (DriftEnd) at ($(TxNode) + (-1.3, 1.3)$); 
        \draw[vector, red, very thick] (TxNode) -- (DriftEnd) node[anchor=south east] {Drift $\mathbf{v}$};
        
        \coordinate (TxExt) at ($(O)!1.4!(TxNode)$); 
        \draw[dashed, gray, thin] (TxNode) -- (TxExt);
        
        \coordinate (A) at ($(TxNode)!0.6cm!(TxExt)$); 
        \coordinate (B) at ($(TxNode)!0.6cm!(DriftEnd)$); 
        \draw[thin, darkgray] (A) to[bend right=45] node[midway, above, font=\footnotesize] {$\psi$} (B);
        
        \draw[orange, thin, ->] (TxNode) 
            -- ++(-0.2, 0.4) -- ++(-0.15, -0.2) -- ++(-0.3, 0.3) -- ++(-0.1, -0.4)  
            -- ++(-0.4, 0.1) -- ++(-0.1, 0.3) -- ++(-0.3, -0.2) -- ++(-0.2, 0.4)    
            -- ++(-0.4, -0.1) -- ++(-0.1, -0.5) -- ($(O)+(55:1.2)$); 
        \node[orange!80!black, font=\footnotesize, align=center] at (3.0, 3.8) {Particle\\Trajectory};      
    \end{tikzpicture}
    }
    \caption{System model of the 3-D molecular communication channel. 
    A point transmitter is located at $\mathbf{x}_0$, and a fully absorbing
    spherical receiver of radius $r$ is at the origin. The medium is subject
    to a uniform drift $\mathbf{v}$ forming an angle
    $\psi=\angle(\mathbf{v},\mathbf{x}_0)$.}
    \label{fig:system_model}
\end{figure}

\section{System Model}\label{sec:system}

We consider an unbounded 3-D fluid environment
$\mathbb{R}^3$.
The transmitter (Tx) is modeled as a point source located at
$\mathbf{x}_0\in\mathbb{R}^3$, and the receiver (Rx) is a \textit{fully-absorbing (FA)}
sphere of radius $r$ centered at the origin, denoted by
$\mathcal{S}_r$, with $|\mathbf{x}_0|>r$.
The system geometry is illustrated in Fig.~\ref{fig:system_model}.

\color{black}
\begin{remark}[Practical relevance of the FA model]
The FA receiver model is a widely adopted transport-level abstraction within MC (see \cite[Fig.~8]{Jamali:2018}). Biochemically, it represents the limiting behavior ($\kappa_f \to \infty$) of high-affinity ligand--receptor binding, where molecules are effectively removed upon contact with the receiver surface; numerical studies indicate that even a modest receptor surface coverage (on the order of $1\%$) suffices to closely approximate the absorption statistics of an ideal FA sphere \cite{Akkaya2015}. 

\end{remark}
\color{black}

Information particles (molecules) are released at time $t=0$.
From a microscopic perspective, the trajectory of a particle
$\mathbf{X}_t$ follows the It\^o stochastic differential equation
\cite{oksendal2003stochastic}, 
\begin{equation}\label{eq:sde}
d\mathbf{X}_t = \mathbf{u}(t)\,dt + \sigma\,d\mathbf{B}_t,
\quad
\mathbf{X}_0=\mathbf{x}_0,
\end{equation}
where $\mathbf{B}_t$ is a standard 3-D Brownian motion,
$D$ is the diffusion coefficient, and $\sigma=\sqrt{2D}$.

While the proposed measure-change framework applies to general drift profiles, this
letter focuses on the practically relevant case of \emph{constant
uniform drift},
\(
\mathbf{u}(t)\equiv\mathbf{v}\in\mathbb{R}^3
\),
forming an angle $\psi=\angle(\mathbf{v},\mathbf{x}_0)$ with the
Tx--Rx axis.
The relative strength of advection and diffusion is characterized by
the P\'eclet number $\mathrm{Pe}=|\mathbf{v}|r/D$.
The first hitting time is defined as
\(
T \triangleq \inf\{t>0:\ |\mathbf{X}_t|=r\},
\)
and the CIR is the probability density
function of $T$, denoted by $f_{\mathrm{3D}}^{(\mathbf{v})}(t)$.

From a macroscopic viewpoint, the corresponding particle concentration
satisfies the advection--diffusion equation with absorbing boundary
conditions.
However, for arbitrary drift directions, direct PDE-based approaches
become analytically intractable, which motivates the stochastic
measure-change formulation adopted in the next section.

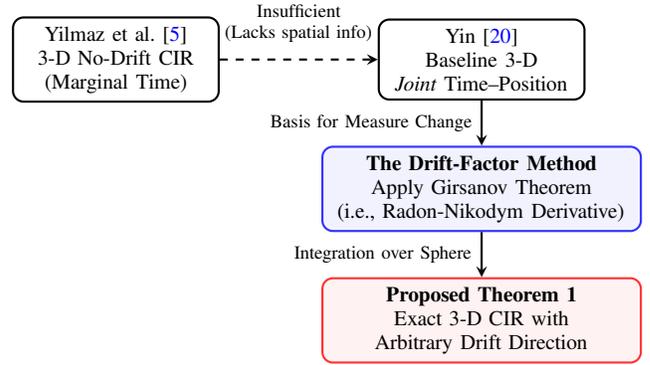
\begin{figure}[t!]
    \centering
    \begin{tikzpicture}[
      node distance=2.0cm,
      auto,
      base/.style={rectangle, draw=black, thick, text centered, rounded corners, minimum height=0.8cm, text width=2.5cm, font=\footnotesize},
      process/.style={rectangle, draw=blue!80, thick, fill=blue!5, text centered, rounded corners, minimum height=0.8cm, text width=4.0cm, font=\footnotesize},
      result/.style={rectangle, draw=red!80, thick, fill=red!5, text centered, rounded corners, minimum height=0.8cm, text width=4.0cm, font=\footnotesize},
      arrow/.style={thick,->,>=stealth}
    ]
      \node [base] (yilmaz) {Yilmaz et al. \cite{Yilmaz:2014} \\ 3-D No-Drift CIR \\ (Marginal Time)};
      \node [base, right=2.1cm of yilmaz] (yin) {Yin \cite{Yin:2009} \\ Baseline 3-D \\ \textit{Joint} Time--Position};
      \node [process, below=0.6cm of yin] (girsanov) {\textbf{The Drift-Factor Method} \\ Apply Girsanov Theorem \\ (i.e., Radon-Nikodym Derivative)};
      \node [result, below=0.6cm of girsanov] (final) {\textbf{Proposed Theorem 1} \\ Exact 3-D CIR with \\ Arbitrary Drift Direction};
      
      \draw [arrow, dashed] (yilmaz) -- node[midway, above, font=\scriptsize, text width=2.5cm, align=center, yshift=0.1cm] {Insufficient \\ (Lacks spatial info)} (yin);
      \draw [arrow] (yin) -- node[left, font=\scriptsize] {Basis for Measure Change} (girsanov);
      \draw [arrow] (girsanov) -- node[left, font=\scriptsize] {Integration over Sphere} (final);
    \end{tikzpicture}
    \caption{Derivation flow based on drift-factor method. We start from Yin's joint time-position distribution \cite{Yin:2009} and apply the Girsanov theorem to derive the 3-D CIR in analytical form.}
    \label{fig:logic_flow}
\end{figure}


\section{Derivation of the Drifted 3-D CIR}\label{sec:derivation}

This section derives the exact CIR for a
spherical absorbing receiver in 3-D space under a uniform
drift with arbitrary direction.
The overall derivation flow is summarized in Fig.~\ref{fig:logic_flow}.
The key idea is to bridge the well-established no-drift case and the
general drifted case through a measure-change argument.

\subsection{Motivation: Why the Joint Distribution Is Necessary}

Classical analytical results for spherical FA receivers without drift typically rely on the marginal first-hitting-time distribution \cite{Yilmaz:2014}, as radial symmetry renders the angular coordinate immaterial. Under drift, however, advection introduces directional bias that breaks this symmetry, so the channel response depends not only on \emph{when} particles arrive but also on \emph{where} they hit the spherical boundary. As a result, the marginal hitting-time distribution alone is insufficient to characterize the drifted CIR. To access this spatial information, we begin from the joint density of the first hitting time $T$ and the hitting location $\mathbf{X}_T$ on $\mathcal{S}_r$, denoted by $f_{\mathrm{3D}}^{(\mathbf{0})}(t,\mathbf{y})$, for which an explicit analytical expression in the no-drift case is available in the probability literature \cite{Yin:2009} and serves as the baseline of our derivation.

\subsection{Drift-Factor Representation via Girsanov Theorem}

To incorporate drift, we employ the Girsanov theorem
\cite{oksendal2003stochastic} to perform a change of measure from a
reference no-drift process (measure $\mathbb{Q}$) to the drifted process
(measure $\mathbb{P}$).
Evaluated at the hitting time, this change of measure introduces a
multiplicative \emph{drift factor}, which analytically relates the
drifted joint density to the no-drift one through
\begin{equation}\label{eq:drift_factor}
f_{\mathrm{3D}}^{(\mathbf{v})}(t,\mathbf{y})
=
\exp\!\left(
\frac{\mathbf{v}\cdot(\mathbf{y}-\mathbf{x}_0)}{\sigma^2}
-
\frac{|\mathbf{v}|^2 t}{2\sigma^2}
\right)
f_{\mathrm{3D}}^{(\mathbf{0})}(t,\mathbf{y}).
\end{equation}

Physically, the drift factor reweights each hitting event according to
the work done by the drift along the particle trajectory.
Hitting locations aligned with the drift direction are exponentially
favored, while those against the drift are suppressed.
This representation isolates the effect of drift into a single,
explicit weighting term (the drift factor) while preserving the analytical structure of
the no-drift joint distribution.

\subsection{Baseline Joint Density for the No-Drift Case}

For a Tx located outside the spherical Rx
(i.e., $|\mathbf{x}_0|>r$), the joint density of the hitting time-location for a 3-D Brownian motion without drift admits
the following series representation, adapted from
\cite[Theorem~3.2]{Yin:2009}, 
\begin{align}\label{eq:yin_joint_3d}
f_{\mathrm{3D}}^{(\mathbf{0})}(t,\mathbf y)
&=
\frac{1}{2\pi}
\frac{r}{|\mathbf x_0|}
\sum_{m=0}^\infty
\Big(m+\tfrac12\Big)
P_m\!\big(\cos\angle(\mathbf x_0,\mathbf y)\big)
\nonumber\\
\times&
\int_0^\infty
\frac{\lambda\,
Z_{m+\frac12}(|\mathbf x_0|/r,\lambda r/\sigma)}
{J_{m+\frac12}^2(\lambda r/\sigma)
+Y_{m+\frac12}^2(\lambda r/\sigma)}
\exp\!\left(-\tfrac12\lambda^2 t\right)
\,d\lambda,
\end{align}
where $P_m(\cdot)$ denotes the Legendre polynomial and
$Z_\nu(a,b)=J_\nu(ab)Y_\nu(b)-J_\nu(b)Y_\nu(ab)$
denotes the standard cross-product of Bessel functions
under normalized radial coordinates.

\subsection{CIR as Surface Marginalization}

The drifted CIR is defined as the marginal distribution of the random
hitting time $T$, obtained by integrating the drifted joint density over
the spherical boundary:
\begin{equation}\label{eq:cir_integral_def}
f_{\mathrm{3D}}^{(\mathbf{v})}(t)
=
\int_{\mathcal{S}_r}
f_{\mathrm{3D}}^{(\mathbf{v})}(t,\mathbf{y})
\,ds(\mathbf{y}).
\end{equation}
Substituting \eqref{eq:drift_factor} into
\eqref{eq:cir_integral_def}, the $\mathbf{y}$-independent term
$\exp\!\left(-|\mathbf{v}|^2 t / 2\sigma^2
-\mathbf{v}\cdot\mathbf{x}_0/\sigma^2\right)$ can be factored out.
The remaining task is to evaluate surface integrals of the form
$e^{\mathbf{v}\cdot\mathbf{y}/\sigma^2}$ multiplied by the angular modes
in \eqref{eq:yin_joint_3d}.

\subsection{Mode-Wise Surface Integration}

To evaluate these integrals, we adapt the following identity from \cite[Lemma~2.5]{Yin:2009}
specialized to the 3-D setting.

\begin{lemma}[3-D Surface Integral Identity]\label{lemma:integral_identity_3d}
For any vector $\mathbf{c}\in\mathbb{R}^3$,
\begin{align}
&\int_{\mathcal{S}_{r}}
e^{\mathbf{c}\cdot \mathbf{y}}
P_m\!\big(\cos\angle(\mathbf{x}_0,\mathbf{y})\big)
\,ds(\mathbf{y}) \nonumber\\
&\qquad=
2(r\pi)^{\frac{3}{2}}
\left(\frac{|\mathbf{c}|}{2}\right)^{-\frac12}
I_{m+\frac12}(|\mathbf{c}|r)\,
P_m\!\big(\cos\angle(\mathbf{c},\mathbf{x}_0)\big),
\end{align}
where $I_\nu(\cdot)$ denotes the modified Bessel function of the first
kind.
\end{lemma}

Applying Lemma~\ref{lemma:integral_identity_3d} with
$\mathbf{c}=\mathbf{v}/\sigma^2$, the angular argument reduces to
$\angle(\mathbf{v},\mathbf{x}_0)=\psi$.
Substituting \eqref{eq:yin_joint_3d} into
\eqref{eq:cir_integral_def}, the CIR decomposes into a summation over
angular eigenmodes $m$, where each term involves a single surface integral
that can be evaluated explicitly using
Lemma~\ref{lemma:integral_identity_3d}.
Collecting all prefactors yields the exact 3-D CIR stated below in
Theorem~\ref{thm:main}.

\begin{theorem}[Exact 3-D CIR under uniform drift]\label{thm:main}
In $\mathbb{R}^3$, the CIR for a spherical
absorbing receiver of radius $r$ under a uniform drift $\mathbf{v}$ is,
\begin{align}\label{eq:main_result}
f_{\mathrm{3D}}^{(\mathbf{v})}(t)
=&- \exp\!\left(
-\frac{\mathbf{v}\!\cdot\!\mathbf{x}_{0}}{\sigma^{2}}
-\frac{|\mathbf{v}|^{2}t}{2\sigma^{2}}
\right)
\frac{\sqrt{2}\,\sigma}{\sqrt{\pi}\,|\mathbf{v}|^{1/2}|\mathbf{x}_{0}|^{1/2}}
\nonumber\\
&\times\sum_{m=0}^{\infty}\Big(m+\tfrac12\Big)
I_{m+\frac12}\!\left(\frac{|\mathbf{v}|r}{\sigma^{2}}\right)
P_{m}(\cos\psi)
\nonumber\\
&\times\int_{0}^{\infty}
\frac{\lambda\, Z_{m+\frac12}(|\mathbf{x}_{0}|/r,\lambda r/\sigma)}
{J_{m+\frac12}^{2}(\lambda r/\sigma)+Y_{m+\frac12}^{2}(\lambda r/\sigma)}
e^{-\frac{1}{2}\lambda^{2}t}\,d\lambda,
\end{align}
where $\psi=\angle(\mathbf{v},\mathbf{x}_{0})$, $\sigma^2=2D$,
$I_\nu(\cdot)$ is the modified Bessel function of the first kind, and
$P_m(\cdot)$ is the Legendre polynomial.
\end{theorem}

\color{black}
\begin{remark}[Numerical Convergence and Evaluation]
\label{rmk:2}
The series in \eqref{eq:main_result} converges rapidly for $t > 0$ because the modified Bessel function $I_{m+1/2}(z)$ decays at a factorial rate (specifically, $I_\nu(z) \sim \frac{(z/2)^\nu}{\Gamma(\nu+1)}$ for large $\nu$) as the order $m$ increases, causing higher-order modes to vanish rapidly under moderate drift regimes. In practice, selecting a truncation order $M$ such that the contribution of omitted terms is negligible relative to the numerical integration tolerance (e.g., set to $10^{-6}$) ensures robust numerical stability. While a formal analytical error bound is reserved for future work, this exact representation serves as the basis for the numerical evaluation and performance characterization in Section~\ref{sec:numerical}.
\end{remark}
\color{black}


\begin{figure*}[!t]
    \centering
    \begin{subfigure}[b]{0.32\textwidth}
        \centering
        \includegraphics[width=\linewidth]{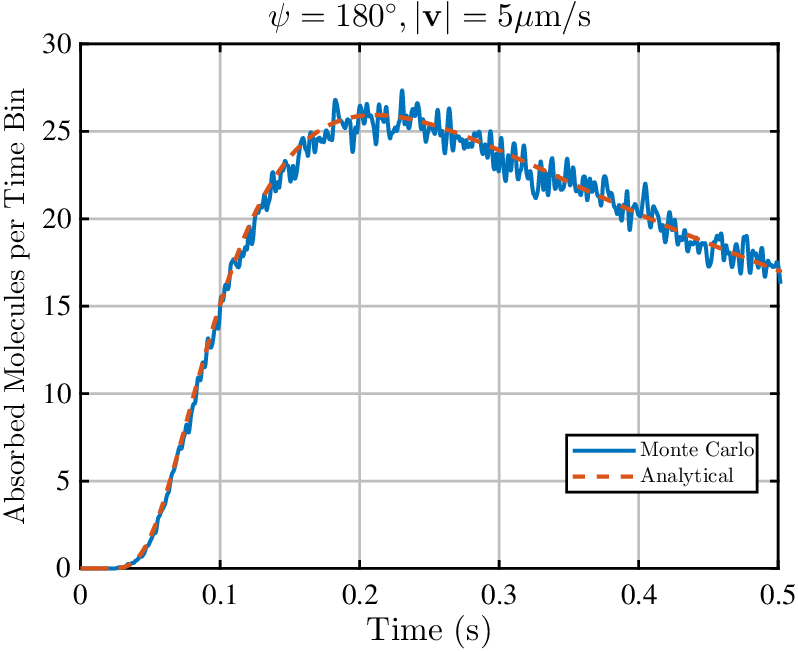}
        \caption{Drift towards Rx, lower speed}
        \label{fig:v5_180}
    \end{subfigure}
    \hfill
    \begin{subfigure}[b]{0.32\textwidth}
        \centering
        \includegraphics[width=\linewidth]{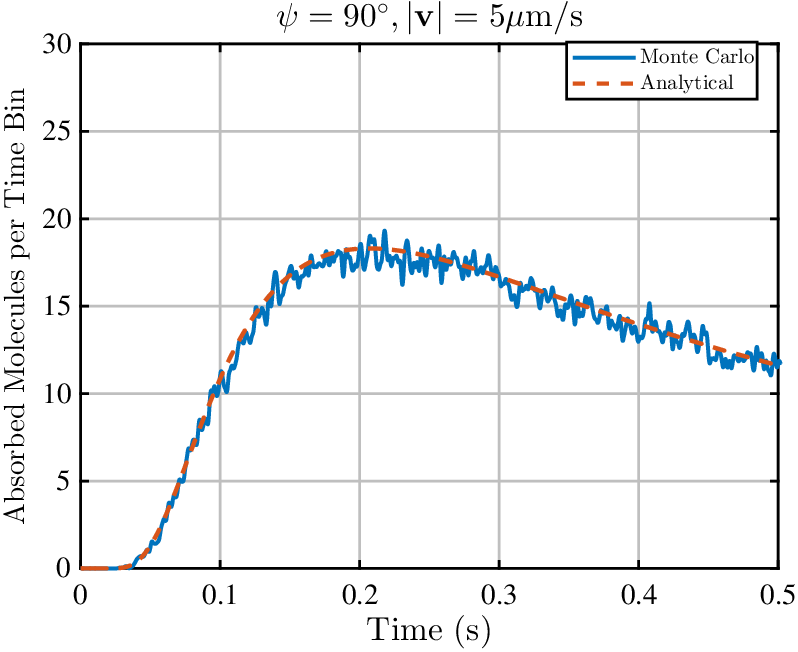}
        \caption{Transverse drift, lower speed}
        \label{fig:v5_90}
    \end{subfigure}
    \hfill
    \begin{subfigure}[b]{0.32\textwidth}
        \centering
        \includegraphics[width=\linewidth]{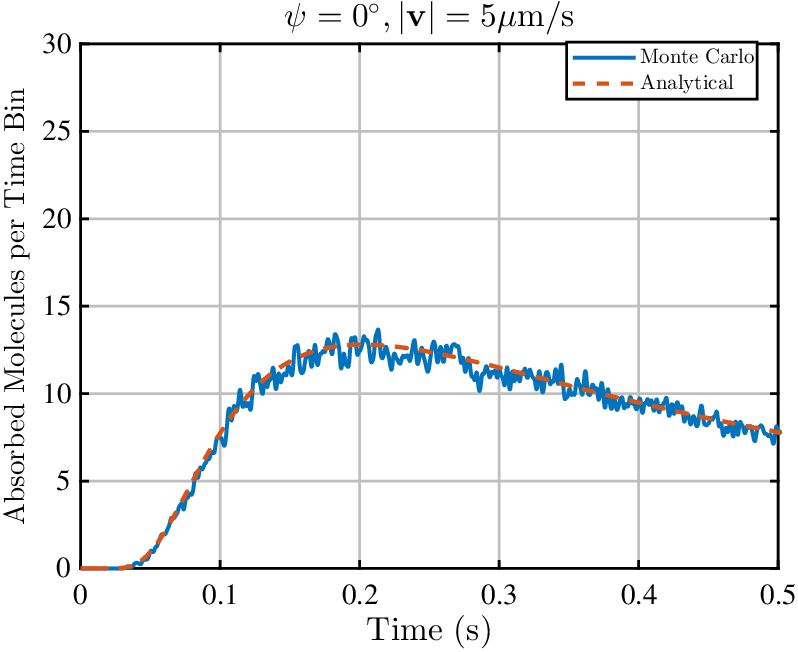}
        \caption{Drift away from Rx, lower speed}
        \label{fig:v5_0}
    \end{subfigure}
    \vspace{0.1cm} 
    \begin{subfigure}[b]{0.32\textwidth}
        \centering
        \includegraphics[width=\linewidth]{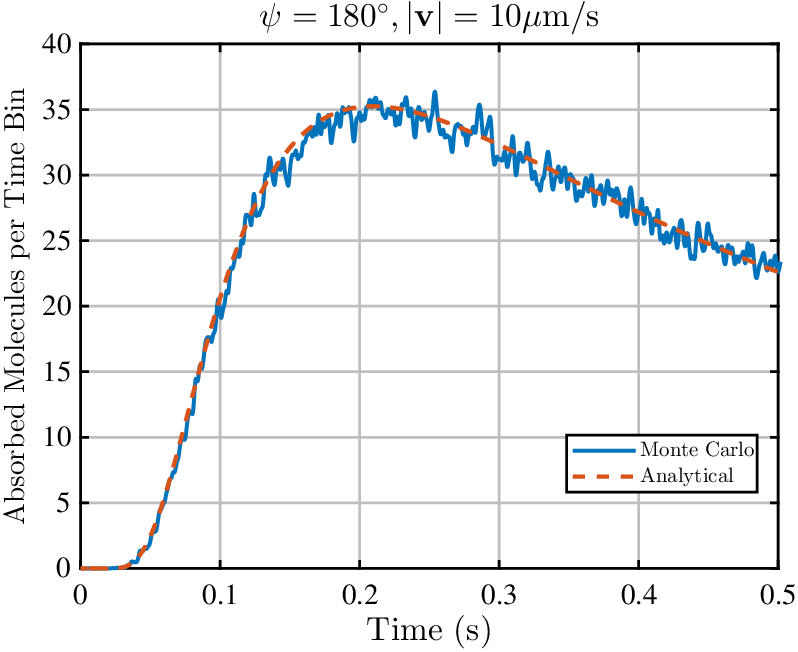}
        \caption{Drift towards Rx, higher speed}
        \label{fig:v10_180}
    \end{subfigure}
    \hfill
    \begin{subfigure}[b]{0.32\textwidth}
        \centering
        \includegraphics[width=\linewidth]{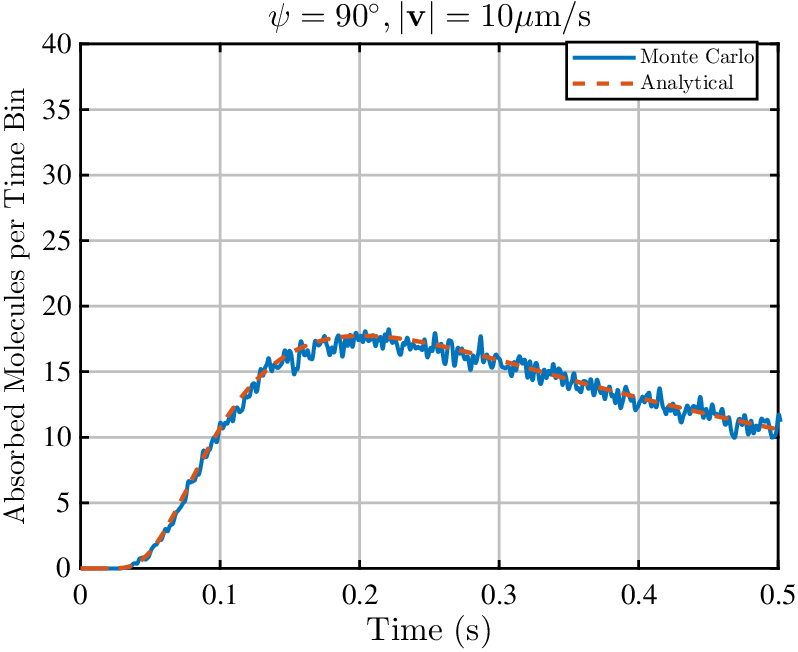}
        \caption{Transverse drift, higher speed}
        \label{fig:v10_90}
    \end{subfigure}
    \hfill
    \begin{subfigure}[b]{0.32\textwidth}
        \centering
        \includegraphics[width=\linewidth]{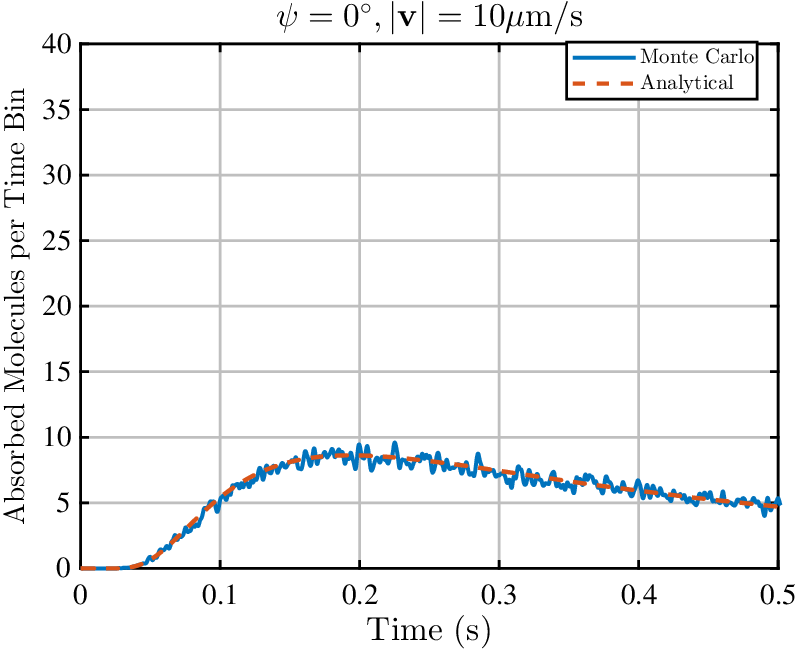}
        \caption{Drift away from Rx, higher speed}
        \label{fig:v10_0}
    \end{subfigure}
    \vspace{-0.2cm}
    \caption{Validation of analytical CIR against Monte Carlo simulations.
    The analytical curves correspond to $N_{\mathrm{tx}} f_{\mathrm{3D}}^{(\mathbf{v})}(t)\,\Delta t$,
    i.e., the expected number of absorbed molecules within a fixed time bin $\Delta t = 5\times10^{-5}\,\mathrm{s}$.
    \textbf{Top Row:} Moderate drift speed ($|\mathbf{v}|=5 \, \mu\mathrm{m/s}$).
    \textbf{Bottom Row:} Higher drift speed ($|\mathbf{v}|=10 \, \mu\mathrm{m/s}$).
    Columns correspond to positive ($\psi=180^\circ$), transverse ($\psi=90^\circ$),
    and negative ($\psi=0^\circ$) drift directions.}
    \label{fig:sim_results}
    \vspace{-0.4cm}
\end{figure*}

\section{Numerical Results and Discussion}
\label{sec:numerical}

We validate the derived analytical CIR against particle-based Monte Carlo simulations and illustrate how the formulation enables efficient characterization of two key channel metrics: pulse amplitude and peak time. All simulations are conducted in MATLAB. Consistent with established MC literature \cite{Yilmaz:2014}, the default parameters are $D = 80 \, \mu\mathrm{m}^2/\mathrm{s}$, receiver radius $r = 10 \, \mu\mathrm{m}$, and initial distance $|\mathbf{x}_0| = 20 \, \mu\mathrm{m}$. Two representative drift magnitudes are considered: $|\mathbf{v}| = 5$ and $10 \, \mu\mathrm{m/s}$.

For the Monte Carlo validation in Fig.~\ref{fig:sim_results}, the analytical CIR is discretized using a fixed time bin $\Delta t$ to match the particle-based implementation. In Figs.~\ref{fig:peak_vs_v} and~\ref{fig:peak_vs_r}, the peak time $t_{\mathrm{peak}}$ is obtained by maximizing the continuous-time analytical CIR, and the corresponding peak amplitude is reported as $N_{\mathrm{tx}} f^{(\mathbf{v})}_{\mathrm{3D}}(t_{\mathrm{peak}})\Delta t$, using the same $\Delta t$ to ensure consistent scaling.


\subsection{Validation against Monte Carlo Simulations}

Fig.~\ref{fig:sim_results} compares the analytical CIR
(Eq.~\eqref{eq:main_result}, truncated at $m=30$) with particle-based
Monte Carlo simulations using $N_{\mathrm{tx}}=10^6$ molecules, for
drift magnitudes $|\mathbf{v}| \in \{5,10\}\,\mu\mathrm{m/s}$ and drift
angles $\psi \in \{0^\circ,90^\circ,180^\circ\}$.
Excellent agreement is observed across all considered scenarios.

As shown in Fig.~\ref{fig:sim_results}, increasing drift aligned with the
Tx--Rx axis sharpens the received pulse, whereas drift
directed away from the receiver suppresses the peak and elongates the
tail.
These trends are accurately captured by the analytical CIR.
Any residual discrepancies are attributable to finite-particle
fluctuations inherent in Monte Carlo simulations.


\color{black}
\subsection{Convergence of the Analytical Series}
To evaluate \eqref{eq:main_result}, the angular series is truncated at order $M$. The convergence is driven by the Gaussian-like decay $e^{-\lambda^2 t/2}$ and the factorial suppression of higher-order Legendre modes. Empirical results demonstrate that selecting $M=30$ ensures the truncation error is negligible relative to standard numerical tolerances. Although pronounced angular anisotropy (e.g., under strong drift or as $t \to 0$) may necessitate a larger $M$ to resolve fine-grained spatial features, we found that $M=30$ provides a near-optimal balance between computational efficiency and analytical precision for the performance characterization in Section~\ref{sec:numerical}.
\color{black}

\subsection{Characterization of Peak Metrics}

Unlike computationally expensive Monte Carlo methods that require extensive averaging to suppress stochastic fluctuations, the proposed exact CIR provides a smooth objective function, enabling highly efficient, noise-free numerical maximization for peak metric extraction.

\begin{figure}[t]
    \centering
    \includegraphics[width=0.9\linewidth]{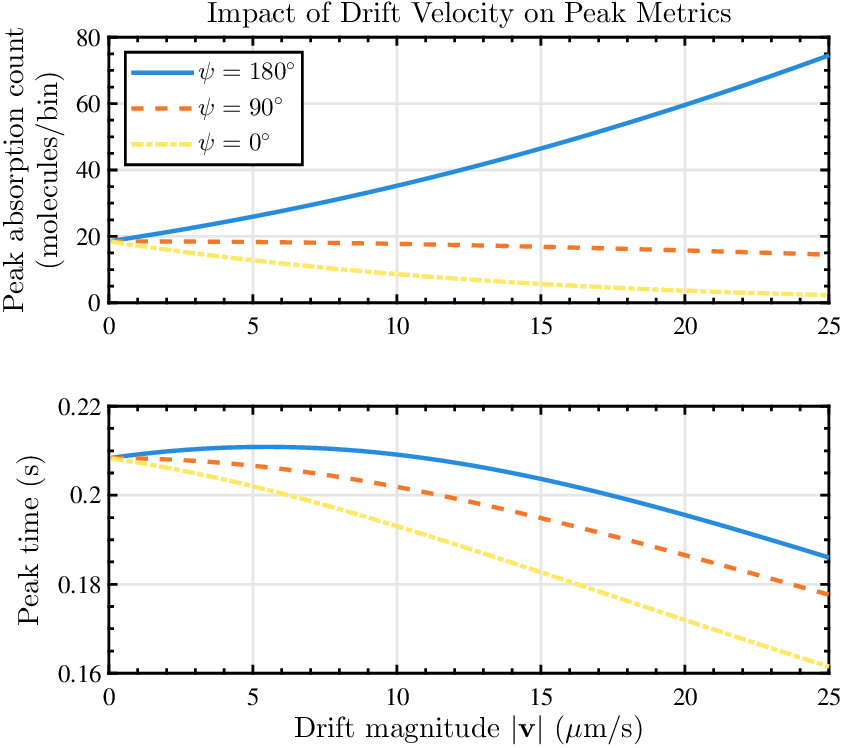}
    \caption{Impact of drift magnitude $|\mathbf{v}|$ on peak metrics for a fixed receiver radius $r=10~\mu\mathrm{m}$. 
    \textbf{Top:} peak absorption count per time bin, computed as 
    $N_{\mathrm{tx}} f^{(\mathbf{v})}_{\mathrm{3D}}(t_{\mathrm{peak}})\Delta t$ with 
    $\Delta t = 5\times10^{-5}\,\mathrm{s}$ (same bin width as in Fig.~3 for consistent scaling).
    \textbf{Bottom:} corresponding peak time $t_{\mathrm{peak}}$, obtained by maximizing the continuous-time analytical CIR and thus independent of $\Delta t$.
    Positive ($\psi=180^\circ$), transverse ($\psi=90^\circ$), and negative ($\psi=0^\circ$) drift directions are shown.
    }
    \label{fig:peak_vs_v}
\end{figure}

Fig.~\ref{fig:peak_vs_v} reports peak metrics versus drift magnitude.
The peak absorption count is reported as the expected number of absorbed
molecules within a fixed time bin $\Delta t$, whereas $t_{\mathrm{peak}}$
is obtained directly from the continuous-time analytical CIR.
As $|\mathbf{v}|$ increases, the peak count grows under positive drift
and is strongly suppressed under negative drift, while $t_{\mathrm{peak}}$
decreases monotonically.

\begin{figure}[t]
    \centering
    \includegraphics[width=0.9\linewidth]{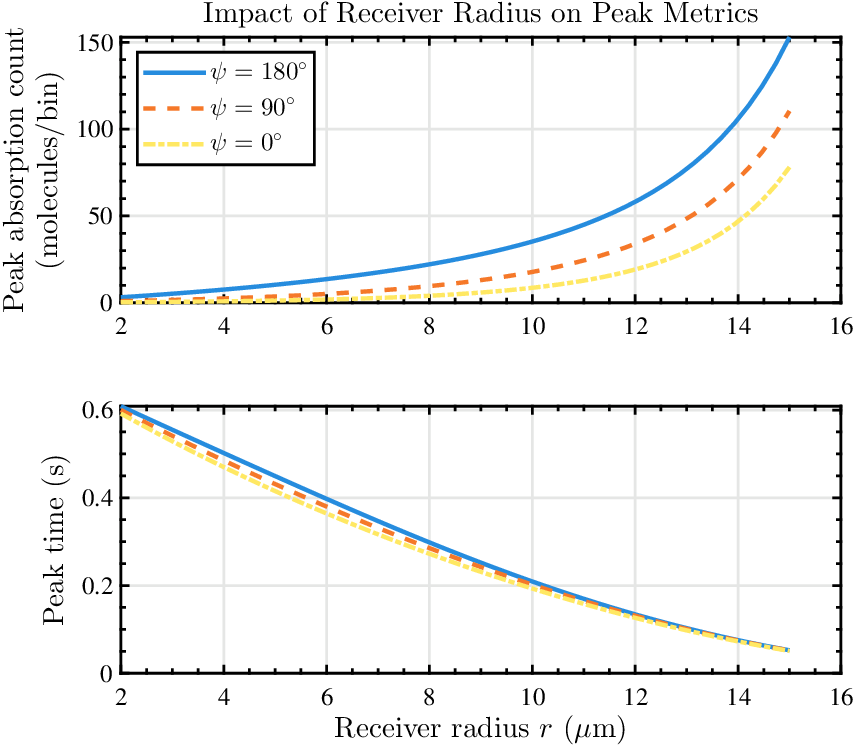}
    \caption{Impact of receiver radius $r$ on peak metrics for a fixed drift magnitude $|\mathbf{v}|=10~\mu\mathrm{m/s}$. 
    \textbf{Top:} peak absorption count per time bin, computed as 
    $N_{\mathrm{tx}} f^{(\mathbf{v})}_{\mathrm{3D}}(t_{\mathrm{peak}})\Delta t$ with 
    $\Delta t = 5\times10^{-5}\,\mathrm{s}$ (same bin width as in Fig.~3 for consistent scaling).
    \textbf{Bottom:} corresponding peak time $t_{\mathrm{peak}}$, obtained by maximizing the continuous-time analytical CIR and thus independent of $\Delta t$.
    Positive ($\psi=180^\circ$), transverse ($\psi=90^\circ$), and negative ($\psi=0^\circ$) drift directions are shown.
    }
    \label{fig:peak_vs_r}
\end{figure}

The trends observed in Fig.~\ref{fig:peak_vs_r} arise from the underlying
geometry of the absorbing receiver rather than from normalization or
parameter choices.
Increasing the receiver radius enlarges the absorbing surface, thereby
enhancing the instantaneous capture probability and leading to a higher
peak absorption count.
At the same time, a larger radius reduces the effective propagation
distance $|\mathbf{x}_0|-r$, which naturally advances the peak time.


\section{Conclusion}
\label{sec:conclusion}

This letter derived an exact analytical 3-D CIR for MC systems with a FA spherical receiver under uniform drift of arbitrary direction. By formulating the problem via joint first-hitting statistics and a Girsanov measure-change argument, we obtained an exact series representation without geometric approximations or dimensional reductions. Verified via particle-based simulations, this formulation clarifies how drift orientation and receiver geometry jointly shape signal enhancement, suppression, and geometric scaling. Furthermore, as a precise and computationally tractable reference model, the proposed CIR \eqref{eq:main_result} enables noise-free peak metric extraction and may facilitate advanced molecular MIMO detection and receiver design \cite{Koo:2016, gursoy2019index, Ahuja:2022}, 
addressing a fundamental challenge in drifted point-to-sphere channel modeling \cite{Yilmaz:2014,Jamali:2018} that has remained open since 2014.


\appendices

\color{black}
\section{Measure-Change Methodological Consistency Check}\label{app:consistency}

This appendix provides a brief \emph{consistency check} for the proposed
measure-change framework (see Fig.~\ref{fig:logic_flow}). 
We consider the 1-D first-hitting-time problem under constant drift to demonstrate that the same measure-change argument recovers the classical IG law from the drift-free L\'evy law.


In one dimension, the absorbing boundary reduces to a deterministic hitting point. Let the initial distance be $\ell>0$ and $\sigma^2=2D$.
The no-drift first-hitting-time density follows the L\'evy law \cite{Srinivas:2012},
\begin{equation}
\label{eq:7}
f_{\mathrm{1D}}^{(0)}(t)
= \frac{\ell}{\sqrt{4\pi Dt^{3}}}
\exp\!\left(-\frac{\ell^{2}}{4Dt}\right), \quad t>0.
\end{equation}
Under constant drift $v$ toward the boundary, the
Girsanov factor at the hitting time reads: 
\(
\exp\!\big(\frac{v\ell}{2D}-\frac{v^{2}t}{4D}\big)
\).
Multiplying \eqref{eq:7} by this factor yields
\begin{equation}
f_{\mathrm{1D}}^{(v)}(t)
=
\frac{\ell}{\sqrt{4\pi Dt^{3}}}
\exp\!\left(-\frac{(\ell-vt)^{2}}{4Dt}\right),
\end{equation}
thereby recovering the classical IG law
\cite{Kadloor:2012,Srinivas:2012}.
\color{black}


\balance
\bibliographystyle{IEEEtran}
\bibliography{exact3D}

\end{document}